\documentstyle[12pt,aasms4,psfig]{article} 
\footnotesep \baselineskip      
\def\lsim{\hbox{\raise.35ex\rlap{$<$}\lower.6ex\hbox{$\sim$}\ }}
\def\ein{{\it Einstein}}

\def\rosat{{\it ROSAT}}

\def\asca{{\it ASCA}}

\def\myarcmin{^\prime\mskip-5mu}
\def\msh{{MSH~11$-$6{\sl 2}}}
\def\g39{{G39.2$-$0.3}}

\def\la{\hbox{\raise.35ex\rlap{$<$}\lower.6ex\hbox{$\sim$}\ }}
\def\ga{\hbox{\raise.35ex\rlap{$>$}\lower.6ex\hbox{$\sim$}\ }}
 
\received{\underline{August 1998}}
\accepted{\underline{November 1998}}
 
\slugcomment{For submission to {\it The Astrophysical Journal}}
 
\begin{document}
\title{AN ASCA STUDY OF THE SUPERNOVA REMNANT G39.2$-$0.3 }
 
\author{ Ilana~M. Harrus \altaffilmark{1,2} and Patrick~O. Slane\altaffilmark{1,3}
 \altaffiltext{1}{ Harvard-Smithsonian Center for Astrophysics,
                   60 Garden Street, Cambridge, MA 02138-1516}
\altaffiltext{2}{ imh@head-cfa.harvard.edu}
\altaffiltext{3}{ slane@cfa.harvard.edu}}
\begin{abstract}
\par\indent 
We present the analysis of archival data from the  {\it Advanced
Satellite for Cosmology and Astrophysics} (\asca)
of the supernova remnant (SNR) \g39. 
\g39\ has been sometimes characterized as a 
shell-like remnant in the literature but our
high-energy imaging and 
spectral analysis show the unambiguous composite nature of the 
remnant. 
We find that part of the contribution to \g39's X-ray emission is 
distinctly non-thermal, best described by a power law with a photon 
index ($2.53^{\rm +0.34}_{\rm -0.27}$). 
The region of emission is consistent with a point source (extension 
consistent  with the point spread function of the detector at the off-axis 
angle of the observation) and is roughly defined by a 
circle of radius $\sim$4$^\prime$.
A second contribution comes from
a thermal component which contributes flux primarily at low energies.
Despite the absence of any pulsed emission detected from the compact source, 
we argue that the emission is most probably due to a rotating 
compact object that is powering the detected synchrotron nebula within the
SNR. \\
\end{abstract}
 
 
\newpage
%
\section{ Introduction }
The standard theory of core collapse of a massive star predicts 
the formation of 
a compact object (neutron star or black hole) as the normal by-product of 
such a supernova explosion. 
Indeed the most famous example of such an 
object is the Crab nebula whose central
pulsar has been extensively studied for nearly three decades, 
studies which have led to the development of 
our understanding of the physics of the interaction of the compact 
object and its surroundings. 
In addition to the 33~msec-pulsar studies, 
the spectral analysis of the Crab Nebula  
and its featureless power-law spectrum has brought to light the 
emission mechanisms at work in the synchrotron nebula. 
Considering that nearly all Type Ib/c or Type II supernovae result 
in the formation of a neutron star,  
it has been suggested that the small number
of confirmed pulsar/SNR associations (less than a dozen) is a
puzzle (Narayan \& Schaudt 1988).  Some authors argue that this 
can be well explained by detection limits of faint pulsars, 
statistical fluctuations and  life time differences between  
SNRs and pulsars (Gaensler \& Johnston 1995).
It remains true, however, that the class of pulsars for which the emission 
beam does not intersect with the Earth is lost to 
pulsed emission searches.
Recent developments in X-ray 
observations have made it possible to circumvent this problem. 
Broadband X-ray observations have provided a tool
to confirm the presence of a compact object by detecting the
associated synchrotron nebula or plerion. This method does not suffer
from the beaming effect bias and gives direct information on the
current energy output of the central
powering object.  With the advance of the high sensitivity broadband
X-ray imaging and spectral instruments onboard \asca, the study of this
class of objects has made great progress.\\

The classification of \g39\ has always been somewhat unclear. 
The remnant was classified as a Crab-like SNR despite the fact that its 
spectral index was more typical of a  shell-type object 
(Caswell et al 1975; Caswell et al 1982) both because the maximum
surface brightness at 1.4GHz is located at the center of the remnant, 
and in view of the large polarization measured across the remnant. 
Polarization is measured both at 1400 and 500 MHz (Caswell et al 1982). 
It is distributed over the remnant with a peak of 10\% polarization in the 
western and central parts and 50\% in the eastern part.\\
Becker \& Helfand (1987) argued that the enhanced emission was 
most probably due to a projection effect as opposed to the 
emission of a plerionic component. In view of the shell-like 
spectral index and the lack of any detected spectral 
variation across the remnant, the authors 
concluded that \g39\ was most probably a shell-like remnant. \\
Another observation of \g39\ made by Patnaik et al (1990) using the 
Very Large Array (VLA), confirmed the typical shell-like integrated 
spectral index 
of the remnant ($0.42\pm0.02$).
\g39\ proved to be more surprising: Anderson \& Rudnick (1993)
found that although the average index of the remnant 
was indeed around the value found 
by Patnaik et al (1990), \g39\ showed evidence of spatial variation of 
the spectral index, from ($0.58^{\rm +0.04}_{\rm -0.05}$) at the edge 
to a Crab-like index of ($0.39^{\rm +0.10}_{\rm -0.05}$) at the center.
In the high-energy band,
\g39\ was observed by the \ein\ Imaging Proportional Counter (IPC), 
but the total observation time was small (2568 sec). 
Although clearly detected, the total 
number of photons (80$\pm$10 photons) was insufficient to differentiate 
between a thermal and a non-thermal spectrum (Becker \& Helfand 1987).
No on-axis pointing of the
object was made with either instrument onboard \rosat\ 
(the Position Sensitive Proportional Counter or the High Resolution 
Imager). \\
In view of neutral-hydrogen absorption 
measurements (Caswell et al 1975) a minimum distance between 7.7 
and 11.3~kpc is generally assumed. 
Becker \& Helfand (1987) estimated that the hydrogen column-density 
toward the remnant is 2$\times$10$^{22}$ atoms~cm$^{-2}$ based on 21cm
absorption measurements with the VLA and concurred on the 
estimate of the distance. \\

Presented here are the results and interpretations of archival 
\asca\ X-ray data on the SNR \g39. 
We present in \S 2 our data extraction methods
and spatial, spectral, and timing analyses. In \S 3 
we discuss the implications of our results.
The final section of the article is a summary of our principal conclusions.

\section{Data Reduction and Analysis}

We retrieved the \asca\ data for \g39\  directly from the HEASARC 
(High Energy Astrophysics Science Archive Research Center) public database. 
Data provided with standard screening process were used for the analysis
(see http://adfwww.gsfc.nasa.gov/asca/processing$_{-}$doc/proc/processing.html
 for more information on that screening process).
No special configuration was applied to the time-recording data bits and 
the time resolution is the nominal one: 0.5~sec in
medium-bit-rate mode and 0.0625~sec in high-bit-rate mode for the GIS 
detector (the SIS is not used for timing analysis).

\subsection{Spatial Analysis}
We have generated exposure-corrected, background-subtracted 
images of the
GIS (GIS 2 and GIS 3 combined) and SIS (SIS 0 and SIS 1 combined) 
data in soft (below 3.0~keV) and hard  (above 3.0~keV) 
energy bands for the 40~ksec observation. 
The background for the spatial analysis was
determined from the weighted average of several nominally blank fields
from high Galactic latitude observations with data selection criteria
matched to those used for the SNR data. Exposure maps were generated from
the off-axis effective-area calibrations, weighted by the appropriate
observation time.  Events from regions of the merged exposure map with
less than 10\% of the maximum exposure were ignored. Merged images of
the source data, background, and exposure were smoothed with a
Gaussian of standard deviation, $\sigma=45^{\prime\prime}$.  We
subtracted smoothed background maps from the data maps and divided by
the corresponding exposure map.  

Fig.~1 shows the results of this procedure for 
both the SIS and the GIS detectors.
The high-energy emission, both in the GIS and the SIS, 
appears to be more centrally enhanced than the low-energy one.
At high energy the maximum emission, identical for both SIS and GIS, 
 is located at 19$^{\rm h}$04$^{\rm m}$4.8$^{\rm s}$, 
05$^\circ$27$^\prime$11.2$^{\prime\prime}$ (J2000).
The center of emission  
is slightly different in the soft and hard band
(we measure a 22$^{\prime\prime}$ offset 
mainly along the North-South axis).  
Considering the off-axis angle of 
the X-ray pointing (the observation was made with an average 
off-axis angle of 7$^\prime$15$^{\prime\prime}$)
and the broad point-spread function of the detector, 
such an apparent offset may be purely associated with systematic 
overcorrection of vignetting effects. 
Similarly, we are unable to definitely establish the point-like or 
extended nature of the high-energy emission. 
The high-energy X-ray emission region is compatible with a point 
source (we used a circle of radius 
$\simeq 4^\prime$ to extract events for both the 
SIS and the GIS).

Fig.~2 shows both a 20cm and a 6cm polarized-intensity radio map (data from
Patnaik et al 1990) superimposed with the SIS contours in the
  high-energy band (from 3.0 to 10.0~keV). 
We note the strong positional correlation between the 
maximum of the polarized-emission and 
the center of the high-energy X-ray profile. 

\subsection{Spectral Analysis}

We extracted X-ray spectra from circular regions centered at 
19$^{\rm h}$04$^{\rm m}$2.1$^{\rm s}$,
05$^\circ$26$^\prime$54$^{\prime\prime}$ (J2000) for both GIS and SIS, 
using a radius of 4$^\prime$17$^{\prime\prime}$ in the GIS 
and only 4$^\prime$ in the SIS because of the CCD-chip boundaries.\\
After the standard data selection criteria 
were applied, the spatial cuts described above lead to combined
counts rates of 0.057$\pm$0.003 cnt sec$^{\rm -1}$ and 
0.068$\pm$0.002 cnt sec$^{\rm -1}$ for the SIS and the GIS respectively. \\

 We first fitted the SIS and GIS separately and then jointly after making sure
 that the combined fits were compatible with the separated fit. 
 All results described below were obtained with a simultaneous fit of 
all four sets of data. 
As seen in the spatial analysis, the emission from the remnant is 
localized and centrally peaked, so it was possible to extract background 
spectra from the same field of view both for the SIS and the GIS.
This method has the advantage of eliminating a possible high-energy 
contamination from the Galactic plane.

\par 
The models applied to fit the emission profile of the remnant reflect the 
physics of the phenomena
which could be contributing to the X-ray output of the remnant. 
If the X-ray emission is coming from the interaction of a blast wave 
with the 
interstellar medium (ISM) the emission is expected to be thermal.
Several models are used in X-ray astrophysics to described such plasmas. 
In the first part of our analysis, we have modeled the thermal emission 
with a collisional equilibrium ionization (CEI) model 
(the so-called ``mekal'' model - Mewe, Gronenschild \& van den Oord 1985; 
Mewe, Lemen \& van den Oord 1986; Kaastra 1992)
which is available in the version 10.00 of XSPEC, the 
X-ray spectral analysis package used throughout this analysis. (The other 
collisional equilibrium model available, Raymond \& Smith 1977, leads to 
similar results).
Unless explicitly stated,  we have kept the elemental abundances at
their nominal values as defined by Anders \& Grevesse (1989).\\
 
 Absorption along the line of sight was taken into account 
 with an equivalent
 column density of hydrogen, $N_{\rm H}$, using the cross-sections and
 abundances from Morrison \& McCammon (1983). 
 In all of the following analysis, the gain offsets of both GIS detectors 
 are linked together and allowed to vary freely.
 A gain shift around -3\% is 
 measured in both detectors, consistent with the results from the 
  calibration data analysis done by the \asca-GIS  team 
  (see http://heasarc.gsfc.nasa.gov/docs/asca/cal$_{-}$probs.html for more
 information on calibration issues).
  We link the normalizations of SIS 0 and SIS 1 spectra (and similarly
  for GIS 2 and GIS 3) while allowing for 
  small variations between the SIS and GIS. We checked that the 
  results for each detector (SIS and GIS) are consistent 
 with each other at the 1$\sigma$ level, considering that due to the 
 CCD boundaries, some flux is consistently missed in 
 the SIS fit normalization.
  As a result, all fluxes quoted hereafter are derived from the GIS results.

  We fitted a pure thermal (CEI model) 
description of the emission and obtained 
  marginally good fit with a $\chi^2$ of 422.3 (and a 
   reduced $\chi^2_{\rm r}= 1.77$). 
  The other single model we tried on the four data sets, is a 
  pure non-thermal model. In this case, we found a worse  
  $\chi^2$ of 492.0 with an associated $\chi^2_{\rm r}$=2.07. 
 In both cases the associated column density is above 
 3$\times$10$^{\rm 22}$ atoms~cm$^{\rm -2}$ consistent with the 
 derived values from the IPC analysis (Becker \& Helfand 1987). 
 A combined fit including both thermal and non-thermal 
  contributions leads to a drop of more than 200 
 in $\chi^2$ (to a value of 241.6) and a $\chi^2_{\rm r}$=1.02. 
 The hydrogen column density is high,(4.70$\pm$0.30)$\times$10$^{\rm 22}$
atoms~cm$^{\rm -2}$,  as expected from the distance estimate 
and the location of the remnant, and the
 temperature is $kT$= 0.61$\pm$0.10~keV. The 
 power-law fit leads to a photon index of 2.08$^{\rm +0.46}_{\rm -0.25}$. 
We also tried to fit the high-energy part of the spectrum with a 
 bremsstrahlung model. In this case, the temperature associated 
is high (4.0$^{\rm +2.0}_{\rm -1.0}$ keV). 
The column density is consistent to the one found with the previous model
(4.71$^{\rm +0.35}_{\rm -0.33}\times$10$^{\rm 22}$ atoms~cm$^{\rm -2}$) 
and so is the temperature of the low-energy component 
($kT$= 0.51$^{\rm +0.08}_{\rm -0.06}$~keV). 
The $\chi^2$ is worse ($\chi^2_{\rm r}$=1.16) 
than the one found with the previous power-law/thermal mixed model
and we favor the power-law
interpretation of the high-energy component.\\
We note that the column density is high relative to that determined from  
HI absorption measurements. Such discrepancies are not uncommon: 
   Arabadjis \& Bregman (1998) show that,  at low galactic latitudes, 
 X-ray derived values are roughly double that of the ones deduced 
from 21~cm measurements, presumably indicating the presence of additional 
molecular material. 

We have also examined the effects of non-equilibrium ionization (NEI) 
on our previous 
results. Non equilibrium effects arise when the  
ions are not instantaneously ionized to their equilibrium temperature  behind 
the shock front. Rather, the timescale for attaining full equilibrium
ionization is comparable to the remnant dynamical timescale. To incorporate
these effects into the model of SNR spectral emissivity, we use here 
the matrix
inversion method developed by Hughes \& Helfand (1985). The ionization state 
depends on the product of electron density and age and we define the 
ionization timescale as  $\tau_i \equiv n_et$. 
As in the CEI model, we have kept all elements to their
nominal abundances (Anders \& Grevesse 1989).
From the results obtained in the previous analysis, we can estimate an upper
limit for the expected value 
of the ionization timescale. We find an upper limit for 
 $n_et$ between 6780 and 15606 yrs~cm$^{\rm -3}$. 
As in the CEI analysis, 
a single model fit cannot account
for the total spectra. 
In the case of a single 
NEI model, we find  $\chi^2$= 287.80  
for 232 degrees of freedom larger than the best
 $\chi^2$ of the complete analysis but much 
 better than the one obtained in the equilibrium case alone. 
Upon adding a power-law component we
find a $\chi^2$= 238.11 (for 235 degrees of freedom). The associated 
column density is 
 (4.65$^{\rm +0.11}_{\rm -0.26}$)$\times$10$^{\rm 22}$ atoms~cm$^{\rm -2}$. 
The temperature ($kT$= 0.62$^{\rm +0.10}_{\rm -0.03}$~keV)
is in complete agreement with the one found in the CEI model 
and the ionization timescale 
(3810$^{\rm +2095}_{\rm -1510}$~yrs~cm$^{\rm -3}$) 
is consistent with the derived upper limit given above. 
The power-law model gives a slightly steeper 
photon index of 2.53$^{\rm +0.34}_{\rm -0.27}$.	
This good agreement between the equilibrium and non-equilibrium models argues
against a very young remnant for which non-equilibrium effects would be very
strong, a fact that is confirmed by the age estimate derived in the following
section. 
We have searched for variation in elemental abundances (in particular 
for Magnesium, Silicon and Sulfur) and found none. 
 Fig.~3 shows the data and the best-fit power-law plus NEI thermal
 model. The thermal component is plotted separately in the insert to
 the figure. 
 It is clear from the figure alone that the thermal contribution 
 is prevalent at low energy while it is contributing only a small amount
  above 3.0~keV. 
 We argue that the spectral data support the interpretation that the 
X-ray emission from \g39\ consists of non-thermal synchrotron emission 
from a central plerion in addition to a thermal
component arising from the interaction of the blast wave with the ISM. 
Results from the spectral fits are tabulated in Table 1. \\

\subsection{Timing Analysis} 

As noted in our study of \msh\ (Harrus et al 1998) 
it is often the case that 
 no pulse is detected from 
 SNRs which present unambiguous evidence for a compact object; 
 beaming effects, for
 example, may result in a cone of emission which does not intersect the
 Earth. This beaming effect presumably accounts for the relatively low
 number of SNRs  for which an associated radio pulsar has been unambiguously
 identified. 
In our timing analysis of \g39, we have extracted data from 
the central part of the emission and restricted
ourselves to energies  above 3 keV in order to suppress the contribution from
the thermal spectral component associated with the remnant.
We searched for periodicity 
in intervals of 0.0625 sec (the best time resolution 
of our data) up to 20 sec. No significant deviation from a random 
 distribution is found. 
To insure the validity of our method, we have extracted from the HEASARC
archives the data for an SNR in which a similar analysis lead to the 
discovery of a pulse emission. 
Identical cuts (both in position and in energy) are 
applied to  Kes~73 data and the same search leads to the 
 detection (at more than 10$\sigma$  level) of the pulsar located at
the center of this remnant (Vasisht \& Gotthelf 1997). 
On the total 2349 events extracted for the \g39\ timing analysis, 
we estimate a
background from the thermal component around 900 events and  we derived 
a pulsed fraction upper limit between 10\% and 25\% (depending
on the pulse shape) at the 90\% confidence level. 

\section{ Discussion}
 Using a distance compatible with the 
minimum distance to \g39\ between 7.7 and 11.3~kpc 
 (Caswell et al 1975), the supernova 
 remnant defines an X-ray emitting volume of
 $V \simeq 1.94\times10^{\rm 59} \,f\,D_{\rm 10}^{\rm 3}$ $\theta_{\rm 4}^{\rm 3}$ cm$^{\rm -3}$,   where $f$ is the volume filling
factor of the emitting gas within the SNR, 
$D_{\rm 10}$ is the distance to the remnant in units of 10~kpc, and
 $\theta_{\rm 4}$ the angular radius in units of 4$\myarcmin$. 
 From this estimate, the combined result from the thermal component 
  (see Table 1),
 and a ratio $n_e/n_{\rm H}$ of 1.08 (extracted from our NEI analysis), 
 we deduce a hydrogen number density of
$n_{\rm H}  = (0.80^{+0.20}_{-0.17})\,D_{\rm 10}^{\rm -1/2}\,\theta_{\rm 4}^{\rm -3/2}\,f^{\rm -1/2}$ cm$^{\rm -3}$.
 The mass of  X-ray emitting plasma is 
$M_{X} = 175\,D_{\rm 10}^{\rm 5/2}\,f^{\rm 1/2}\theta_{\rm 4}^{\rm 3/2}$ $M_\odot$, which for the distance range mentioned above and combined with the errors
from the spectral analysis,  translates to 
  a mass $M_{X}$ between 40 and 300 $M_\odot$. 
  Considering that the distances used are already a lower bound of the 
  possible distance to the remnant, we argue  that this result is consistent 
  with the remnant being in its Sedov-Taylor phase of
  evolution (Taylor 1950; Sedov 1959).
  In this case, an estimate on the age of the remnant can be made 
  using the temperature obtained in our spectral analysis.
  This temperature, $\langle T\rangle$,  is the
  emission-measure-weighted average electron temperature, which is
  proportional to the shock temperature $\langle T\rangle = 1.27~T_S$
  (assuming that electrons and ions have the same temperature). 
  The Sedov age relationship is 
   $t\approx 480\,{\rm yrs}\,(R /1\,{\rm pc})\,(\langle T\rangle /1\,{\rm keV})^{-1/2}\approx$
    $(7100^{\rm +1150}_{\rm -2020})$ yr 
(the errors bars reflect both the 1$\sigma$ errors on the measured 
temperature and the uncertainties on the distance). 
 The preshock ISM number density is 
 $n_{\rm 0}= \rho_0{\rm m_{\rm H}}= (0.24^{\rm +0.07}_{\rm -0.05})\,D_{\rm 10}^{-1/2}\,\theta_{\rm 4}^{\rm -3/2}\,f^{\rm -1/2}$ cm$^{\rm -3}$, which leads to 
  $n_{\rm 0} = (0.49^{\rm +0.14}_{\rm -0.10})\,D_{\rm 10}^{\rm -1/2}\,\theta_{\rm 4}^{\rm -3/2}$ cm$^{\rm -3}$ for a filling factor of 0.25
(as expected in the case of a pure Taylor-Sedov solution).
With the preceding numerical values we estimate the supernova explosion energy
 from the Taylor-Sedov relations to be 
 $E = (3.5^{\rm +1.5}_{\rm -0.8})\times 10^{\rm 50}\,D_{\rm 10}^{\rm 5/2}\,\theta_{\rm 4}^{\rm 3/2}\,f^{\rm -1/2}$  ergs which is within the range of typical values.\\

  Assuming temperature equilibrium between electrons and ions, we can compute
  the pressure at the shock front 
 $P_s \sim$ 1.93~$n_e kT_S \sim (1.29^{\rm +0.61}_{\rm -0.33})\times$10$^{\rm -9}\,D_{\rm 10}^{\rm -1/2}\,f^{\rm -1/2}$ dyne cm$^{\rm -2}$.
For a Taylor-Sedov temperature and density distribution, the central 
pressure is $P_0 \sim 0.31 P_s$ = 8$\times$10$^{\rm -10}$ dyne cm$^{\rm -2}$ 
for a filling factor of 0.25. 
  This value of the internal thermal pressure leads to an equipartition 
magnetic field value 
 $B = \sqrt{8 \pi P_0} = 
 100^{\rm +20}_{\rm -10}\,D_{\rm 10}^{\rm -1/4}\,f^{\rm -1/4}\,\mu$G. \\
  This estimate of the magnetic field allows us to estimate the 
  break frequency $\nu_{\rm B}$, the frequency at which synchrotron losses
  begin to dominate. We have 
  \begin{equation}
 B = 14.6\,(\nu_{\rm B}/10^{\rm 6}~{\rm GHz})^{\rm -1/3}\,(t/10^4~{\rm yrs})^{\rm -2/3} \,\mu{\rm G}.
 \end{equation}
 (Ginzburg \& Syrovatskii 1965 eqn.~5.36). 
  Here $t$ is the age of the nebula for which we use the
 value derived in our study of the thermal emission.  
 For the results mentioned above and including the uncertainties on the 
 distance and the magnetic field, we find a break frequency 
 between 3350 and 1.25$\times$10$^{\rm 4}$ $f^{\rm 3/4}$ GHz 
 (to be compared to the Crab break frequency  
 $\nu_{\rm break}$ = 10$^{\rm 4}$  GHz)(see Table~2).

 The combined existing radio and X-ray data can provide 
  another estimate of the range of possible break frequencies. 
 Using 
  the spectral index (0.39$^{\rm +0.10}_{\rm -0.05}$) and 
 the measured radio flux from the 
 central region of \g39\ (0.20~mJy arcsec$^{\rm -2}$) 
 (Anderson \& Rudnick 1993), we found a range 
  of frequencies ranging from 1.4$\times$10$^{\rm 5}$ to 
   3$\times$10$^{\rm 6}$ GHz. 
 To reconcile the apparent discrepancy between the two ranges, 
one must note that the actual plerion contribution 
to the radio flux is 
smaller than the total observed flux  
(which includes contribution from the shell) and that 
the spectral index is probably overestimated as well. Both values
affect the value of the break frequency range quoted above.  
As an example, we found that for a spectral index of 0.20 and 
a filling factor of 0.5, the two ranges of frequency are overlapping.\\
 
We have argued above that the nonthermal emission is produced by
a synchrotron nebula surrounding a spinning neutron star, or
pulsar, created in the supernova explosion. 
Using the measured X-ray luminosity associated with the non-thermal component,
L$_{X}^\alpha$ = 2.28$\times$10$^{\rm 35}\,D_{\rm 10}^2$ ergs s$^{-1}$ 
(0.2 -- 4.0 keV) along with the empirical
relationship between $L_{X}^\alpha$ and the 
pulsar's spin-down energy $\dot E$ (Seward \& Wang 1988), we infer a
$\dot E\sim 2.4\times~10^{\rm 37}\, D_{\rm 10}^{\rm 1.44}$~ergs s$^{-1}$ 
for the putative pulsar in \g39. 
Using this value, an assumed breaking index of 3, 
and the age deduced from the Sedov analysis, we derive 
an initial spin period range of 24-36~ms for  
surface magnetic field values between $10^{\rm 12}$ 
and 7$\times10^{\rm 12}$G.
From this initial spin period, one can deduce the current-epoch 
period in the range of 25 to 70~ms.

SNRs with similar characteristics have been found with 
\asca. The class includes \msh\ (Harrus, Hughes \& Slane 1998) and Kes 75 
(Blanton \& Helfand 1996) which have similar values for the age
 (depending upon distance), photon index, and  temperature.
In fact among the objects which belong to the class of SNRs 
for which \asca\ has 
revealed a high-energy emitting center, \g39\ appears to be rather
average. 
  
\section{Summary}
  We have presented the results of \asca\ X-ray spectral and spatial 
studies of the SNR \g39, a middle-aged remnant (about 7000 year old) 
 in its Taylor-Sedov phase of evolution.
 In our model, 
 its X-ray emission consists of 1) a thermal emission from the blast wave
 interacting with the ISM and 2) a non-thermal synchrotron emission 
 due to high-energy electrons surrounding the compact object. 
 We can develop an
 acceptable description for the synchrotron nebula with an 
 associated spin-down power of the undetected and central neutron star
 of  $\dot E = 2.4\times 10^{\rm 37}{\rm\ erg\ s}^{-1}$. \\

\par 

This research has made use of data obtained from the High Energy
Astrophysics Science Archive Research Center (HEASARC), provided by
NASA's Goddard Space Flight Center.
This research was supported in part by NASA under grants NAG5-3486 and
contract NAS8-39073. 
The authors thank Alok Patnaik for providing the radio maps used to 
generate Fig 2 and Lawrence Rudnick for his help. 
Many thanks to  Fred Seward, Paul Plucinsky, and Richard Edgar 
for suggestions and comments, to Keith Gendreau for his help with 
detector issues and to the referee, Rob Petre, for his helpful 
suggestions.\\
IMH would like to thank Ira Glass for his infinite patience and 
Vivien Marx for her support.\\

\pagebreak

\newpage
\vspace{-1.0in}
\noindent

Fig. 1.$--$ \asca\ X-ray images of the SNR \g39\ at low and high energy (top:
0.5--3.0~keV; bottom: \ga 3.0~keV) for the GIS (left) and the SIS
(right). Contour values are linearly spaced from 30\% to 90\% of the peak
surface brightness in each map. Peak/background values are top: 2.58/0.12;
bottom: 1.72/0.08 for the GIS and top: 2.57/0.12; bottom: 1.74/0.19 for the
SIS, where all values are quoted in units of
$10^{-3}$counts~s$^{-1}$~arcmin$^{-2}$.\\

\noindent
Fig. 2.$--${\it Left}: Radio image at 20cm from Patnaik et al (1990)
superimposed with SIS high-energy band (from 3.0 to 10.0 keV) contours. 
The X-ray contours are 
linearly spaced from 30\% to 90\% of the peak surface brightness as defined
in the Fig. 1. 
2.$--${\it Right}. Polarized intensity map for data at 6cm with the 
same X-ray contours superimposed.\\ 

\noindent
Fig. 3.$--$ GIS and SIS spectra -- both GIS (GIS 2 and GIS 3) and SIS (SIS
0 and SIS 1) spectra have been combined for display purposes only.  The solid curves in the top panels show the joint best-fit NEI thermal plasma plus power-law models. The insert figures show only the thermal component; the bottom panels plot the data/model residuals.
\\

\newpage
\footnotesize

\centerline{\bf Table 1}
\centerline{\bf Spectral Analysis of Power Law plus NEI Thermal$^{\rm\  a}$ Plasma Model}
\vspace{0.5cm}
\centerline{\begin{tabular}{lc} \tableline\tableline \\[-8mm]
\multicolumn{1}{l}{Parameter} &
\multicolumn{1}{c}{Fit results$^{\rm b,c}$} \\[1.5mm] \tableline 
$N_{\rm H}$ (atoms~cm$^{\rm -2}$)& (4.65$^{\rm +0.11}_{\rm -0.26}$)$\times$10$^{\rm 22}$\\[1.5mm] \tableline 
$\alpha_{\rm p}$ &  2.53$^{\rm +0.34}_{\rm -0.27}$  \\[1.5mm] 
 $F_{\rm 1 keV}$ (photons cm$^{\rm -2}$ s$^{\rm -1}$ keV$^{\rm -1}$)& (3.4$^{\rm +2.5}_{\rm -1.7}$)$\times$10$^{\rm -3}$\\[1.5mm]
Flux$^\alpha$ (ergs~cm$^{\rm -2}$~s$^{\rm -1}$)~{\rm ([ 0.2 -- 3.0] keV)} & $1.84\times 10^{-11}$  \\[1.5mm] 
Flux$^\alpha$ (ergs~cm$^{\rm -2}$~s$^{\rm -1}$)~{\rm ([ 3.0 -- 10.0] keV)} & $2.71\times 10^{-12}$  \\[1.5mm] 
Flux$^\alpha$ (ergs~cm$^{\rm -2}$~s$^{\rm -1}$)~{\rm ([ 0.2 -- 4.0] keV)} &$1.92\times 10^{-11}$  \\[1.5mm] \tableline
$kT$ (keV) &   0.62$^{\rm +0.10}_{\rm -0.03}$ \\[1.5mm]
log($n_et$) (cm$^{\rm -3}$~s) &   11.08$^{\rm +0.19}_{\rm -0.22}$ \\[1.5mm]
Normalization$^{\rm d}$(cm$^{\rm -5}$) & (9.5$^{\rm +5.3}_{ \rm -3.3}$)$\times$10$^{\rm 12}$    \\[1.5mm] 
Flux$^{\rm Thermal}$ (ergs~cm$^{\rm -2}$~s$^{\rm -1}$)~{\rm ([ 0.2 -- 3.0] keV)} & $4.64\times 10^{-10}$  \\[1.5mm] 
Flux$^{\rm Thermal}$ (ergs~cm$^{\rm -2}$~s$^{\rm -1}$)~{\rm ([ 3.0 -- 10.0] keV)} & $7.05\times 10^{-13}$  \\[1.5mm] 
Flux$^{\rm Thermal}$ (ergs~cm$^{\rm -2}$~s$^{\rm -1}$)~{\rm ([ 0.2 -- 4.0] keV)} & $4.65\times 10^{-10}$  \\[1.5mm]\tableline 
Gain$_{\tiny \it GIS2}$= Gain$_{\tiny \it GIS3}$ & -3.02$^{\rm +0.17}_{\rm -1.54}$ \% \\[1.5mm]
$\chi^{2}$/$\nu$ & 238.11/235  \\[1.5mm] \tableline
\multicolumn{2}{l} {$^{\rm a}$ Abundances from Anders \& Grevesse (1989)}\\
\multicolumn{2}{l} {$^{\rm b}$ Single-parameter 1~$\sigma$ errors}\\
\multicolumn{2}{l} {$^{\rm c}$ All fluxes quoted are unabsorbed.}\\
\multicolumn{2}{l} {$^{\rm d}$ N=(${n_{\rm H}n_eV \over 4\pi D^2}$)}\\
\end{tabular}}
 
\newpage
\centerline{\bf Table 2}
\centerline{\bf Summary of the results} 
\vspace{0.2cm} 
\centerline{\begin{tabular}{lc} \tableline\tableline \\[-0.8cm]
Parameter                      &   \\[1.5mm] \tableline
$R^{\rm\ a}$(pc)     &  11.64 $D_{\rm 10}\,\theta_{\rm 4}$  \\[1.mm] 
Age (yr)  &   6600 --- 7300 $D_{\rm 10}\,\theta_{\rm 4}$ \\ [1.mm]
$n_{\rm 0}^{\rm\  b}$ (cm$^{-3}$)  & $(0.24^{+0.07}_{-0.05})\,D_{\rm 10}^{-1/2}\,\theta_{\rm 4}^{-3/2}\,f^{-1/2}$\\ [1.mm]
$E_{\rm 0}$ (ergs)& $(0.35^{+0.15}_{-0.08})\times 10^{51}\,D_{10}^{5/2}\,\theta_{\rm 4}^{3/2}\,f^{-1/2}$ \\ [1.mm]
M$_{SU}$ (M$_\odot$)  &  140 --- 220 $D_{10}^{5/2}\,f^{1/2}$ \\[1.mm]
$P_0$ ($10^{-10}$ dyne cm$^{-2}$) & 3.0 --- 5.9 $D_{10}^{-1/2}\,f^{-1/2}$ \\[1.mm]
B ($\mu$G)& 90 --- 120 $D_{10}^{-1/4}\,f^{-1/4}$\\ [1.mm]
$\nu_{\rm break}$ (GHz) & 3900 --- 9000 $D_{10}^{-5/4}\,f^{3/4}$ \\[1.mm]
L$_{X}^\alpha$ (ergs~s$^{\rm -1}$)~{\rm ([ 0.2 -- 4.0] keV)} & $2.3\times 10^{35}$ $D_{\rm 10}^2$ \\[1.5mm] \tableline
\multicolumn{2}{l} {$^{\rm a}$ $D_{10}$ is the distance in units of 10~kpc and $\theta_{\rm 4}$ is the angular radius in units of 4$\myarcmin$.}\\ 
\multicolumn{2}{l} {$^{\rm b}$ $f$ is the volume filling factor}\\ 
\end{tabular}}

\newpage
\pagestyle{empty}
\normalsize
\vspace{-.05in}
\begin{figure}[h]
\vspace{-1.0in}
\centerline{\psfig{file=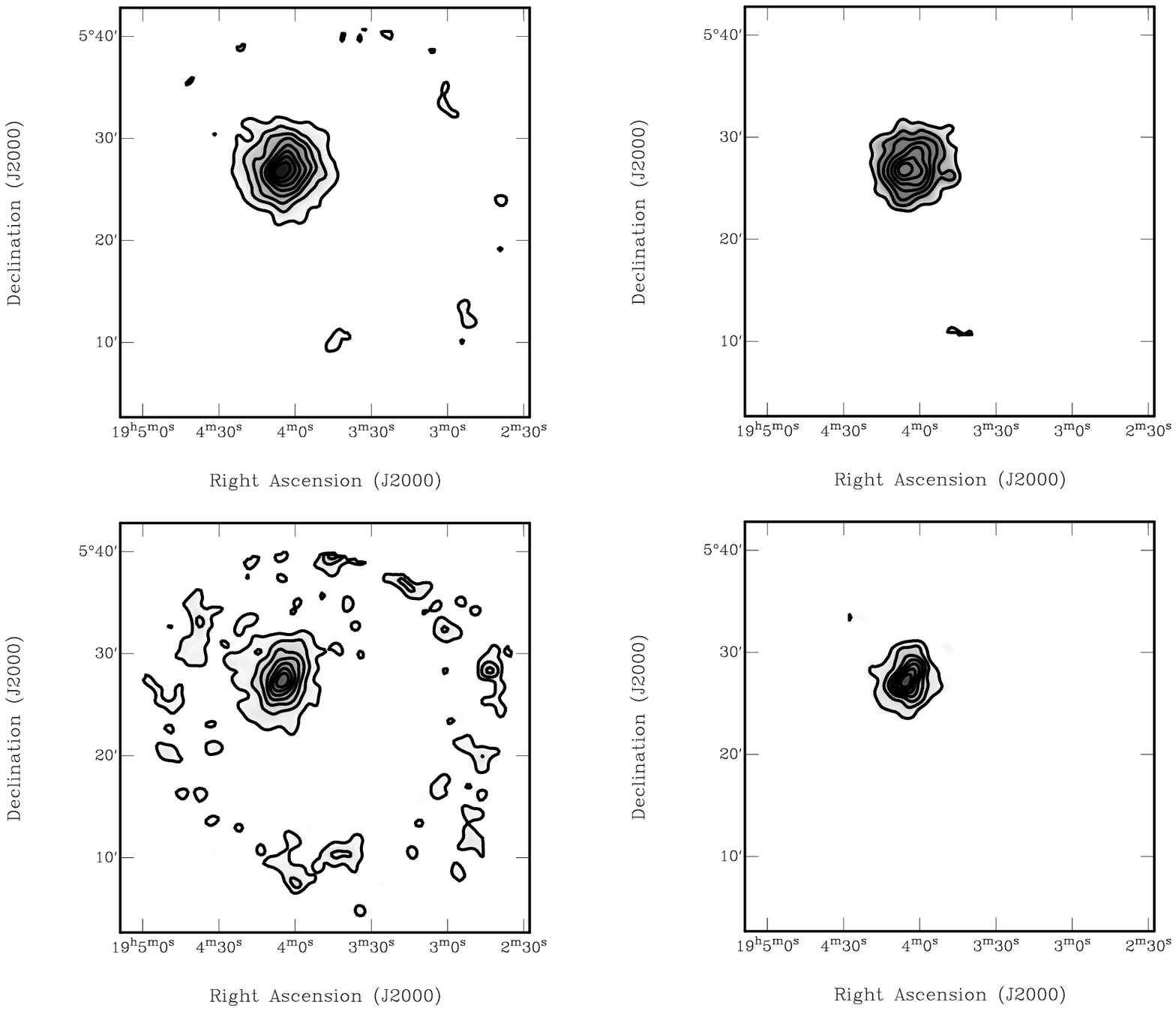,width=6in}}
\end{figure}

\begin{figure}[t]
\vspace{-1.0in}
\centerline{\psfig{file=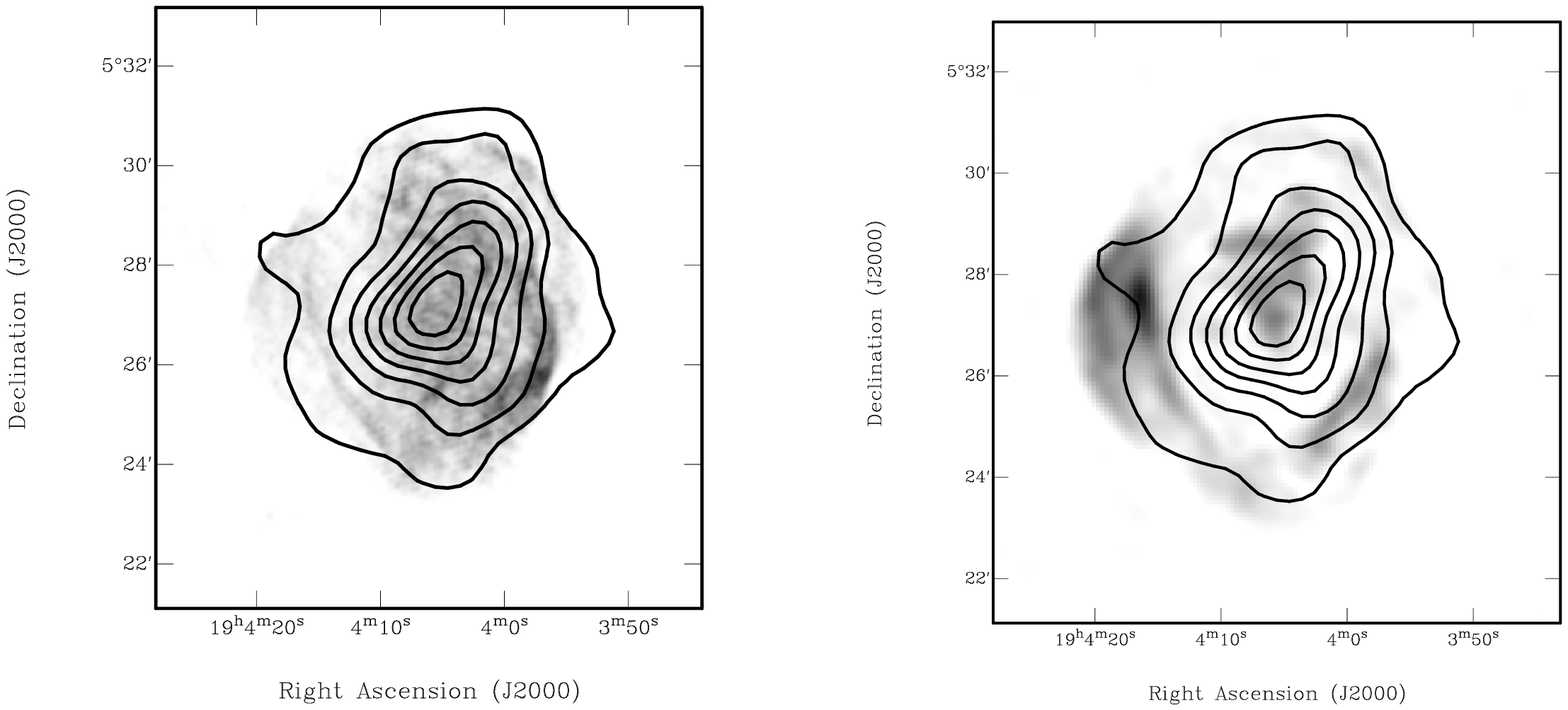,width=7.5in,angle=270}}
\end{figure}
 
\newpage
\begin{figure}[t]
\vspace{-1.5in}
\centerline{\psfig{file=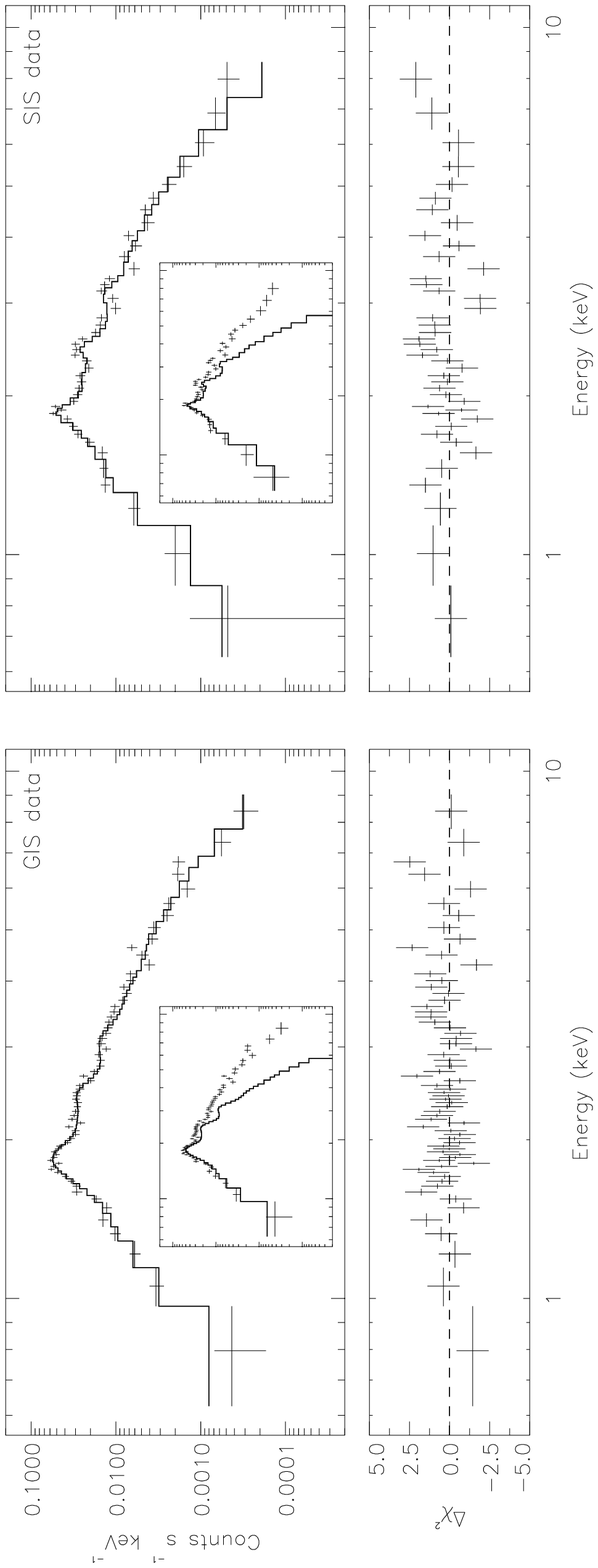,width=6in}}
\end{figure}

\end{document}